\documentclass[showpacs,prd,aps]{revtex4}
\pdfoutput=1

\usepackage{graphicx}

\usepackage{amsmath}
\usepackage{amssymb}
\usepackage{latexsym}
\usepackage{amsfonts}
\usepackage{slashed}
\usepackage{color}
\usepackage{bbold}

\usepackage{hyperref}

\def\hhref#1{\href{http://arxiv.org/abs/#1}{arXiv:#1}} % in bibliography

\newcommand{\ket}[1]{|#1\rangle}

\newcommand{\nn}{\nonumber\\}

\newcommand{\mc}{\mathcal}
\newcommand{\del}{\partial}
\newcommand{\bea}{\begin{eqnarray}}
\newcommand{\ea}{\end{eqnarray}}
\newcommand{\eea}{\end{eqnarray}}

\newcommand{\ta}{\tilde a}
\newcommand{\tb}{\tilde b}

\begin{document}

\title{Electric dipole moment induced by a QCD instanton in an external magnetic field}
%\title{Electric dipole moment of the QCD instanton in an external magnetic field}
%\title{Quarks in an Instanton and Magnetic Field Background}
%\title{QCD Instanton in an External Magnetic Field}

\author{G\"ok\c ce Ba\c sar$^{1,2}$, Gerald~V.~Dunne$^{1}$, and Dmitri~E.~Kharzeev$^{2,3}$}

\affiliation{$^1$ Department of Physics, University of Connecticut, Storrs CT 06269, USA
\\
$^2$Department of Physics,  Stony Brook University, Stony Brook, NY 11794, USA
\\
$^3$Department of Physics,  Brookhaven National Laboratory, Upton, NY 11973, USA}

\begin{abstract}
In the chiral magnetic effect, there is a competition between a strong magnetic field, which tends to project positively charged particles to have spin aligned along the magnetic field, and a chirality imbalance which may be produced locally by a topologically nontrivial gauge field such as an instanton. We study the properties of the Euclidean Dirac equation for a light fermion in the presence of both a constant abelian magnetic field and an $\rm{SU(2)}$ instanton. In particular, we analyze the zero modes analytically in various limits, both on ${\bf R}^4$ and on the four-torus, in order to compare with recent lattice QCD results, and study the implications for the electric dipole moment.
\end{abstract}

%\date{\today}


\maketitle

\date{\today}

\section{Introduction}

Since quarks carry both electric and color charge they couple to both electromagnetic and gluonic gauge fields. A magnetic field introduces a Landau level structure to the fermion spectrum, in which the zero modes have definite spin, aligned along the magnetic field \cite{aharonov-casher,novikov,AlHashimi:2008hr,Giusti:2001ta,Tenjinbayashi:2005sy}. In a gluonic field with nontrivial topological charge the fermion spectrum also has zero modes, with chiralities determined locally by the local topological charge of the gauge field \cite{'tHooft:1976fv,Schwarz:1977az,Kiskis:1977vh,Brown:1977bj,Jackiw:1977pu,Jackiw:1983nv,Shifman:1999mk,rubakov}. The associated instanton transitions in the $\theta$-vacuum can lead to a fluctuating electric dipole moment of the neutron \cite{faccioli}.
In this paper we investigate what happens when a quark experiences both a strong magnetic field and a topologically nontrivial gluonic field, such as an instanton. For a single instanton the fermion spectral problem has a conformal symmetry \cite{Jackiw:1976dw,Chadha:1977mh}, and the zero modes are localized on the instanton, falling off as a power law with Euclidean distance. The conformal symmetry is broken by the introduction of a magnetic field, and now the zero modes develop an asymmetry, falling off in Gaussian form in the plane transverse to the B field, but as a power law in the other two directions. This basic asymmetry is the key to the phenomenon of magnetic catalysis \cite{Gusynin:1995nb} and the chiral magnetic effect \cite{Kharzeev:2004ey,arXiv:0706.1026,arXiv:0711.0950,Fukushima:2008xe,Kharzeev:2009fn}, as sketched in Fig. 1.
\begin{figure}[htb]
\includegraphics[scale=0.3]{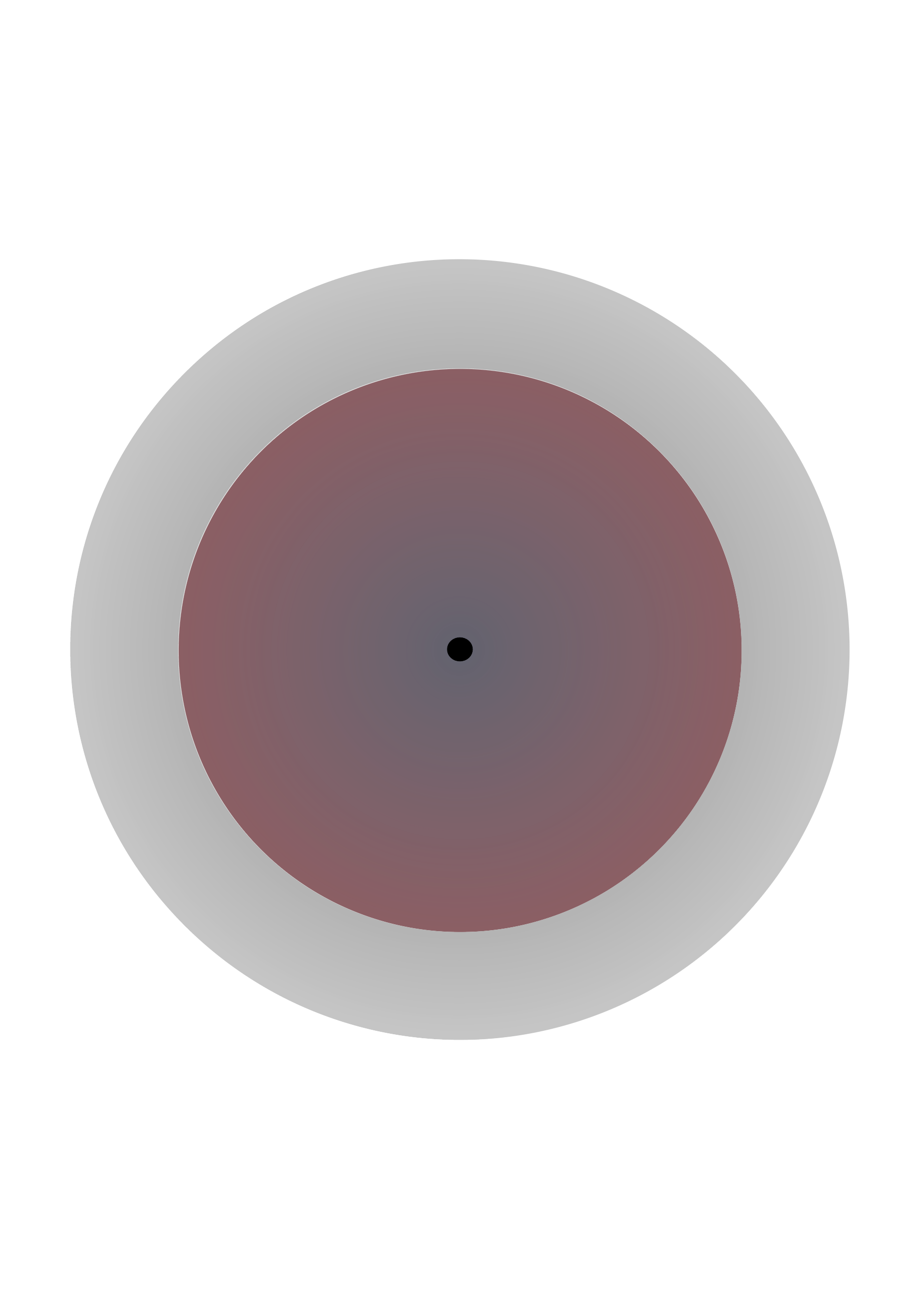}\qquad\qquad 
\includegraphics[scale=0.3]{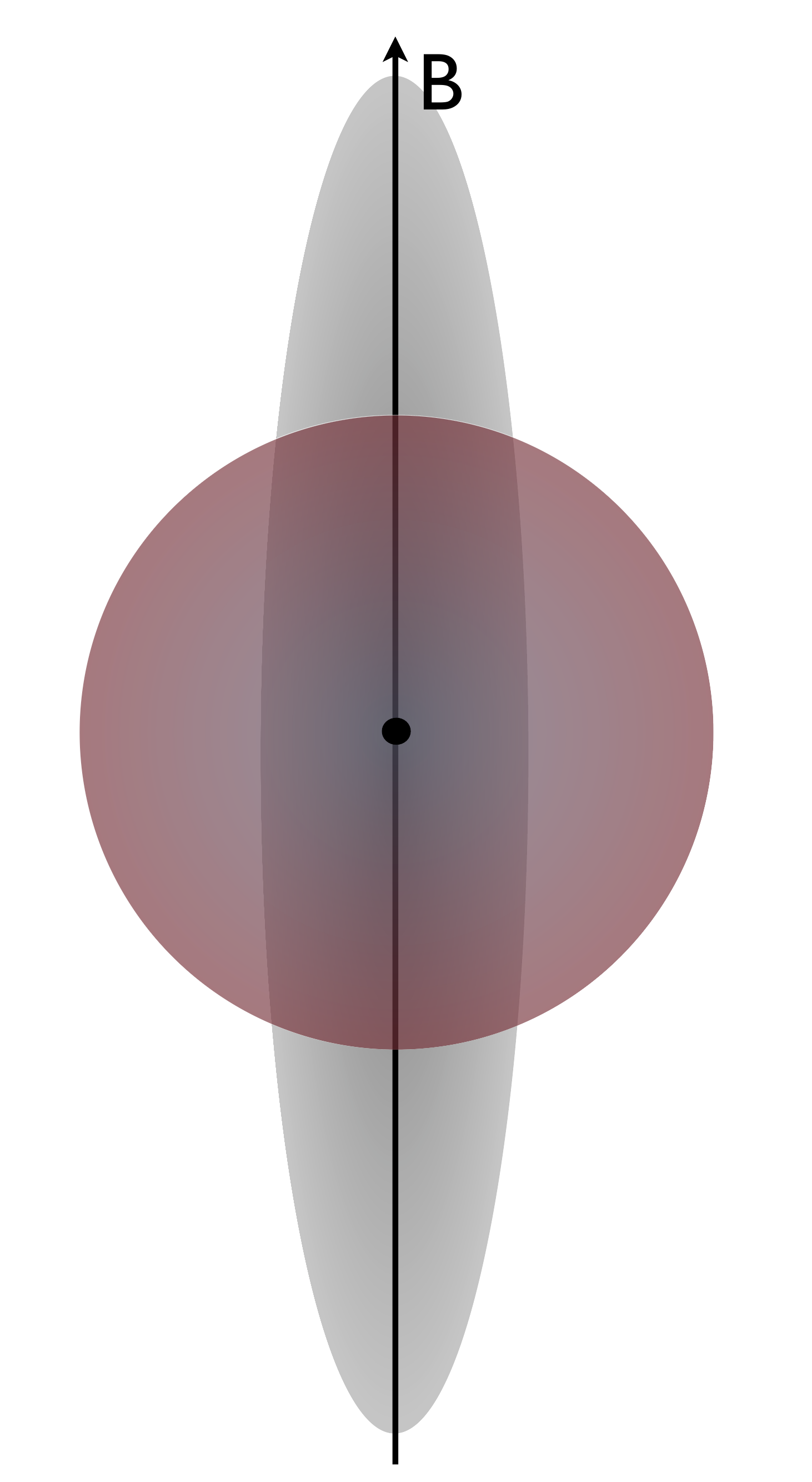}
\caption{A sketch of the topological charge density, $q\propto{\rm tr}\mc F_{\mu\nu}\tilde{\mc F}_{\mu\nu}$, for a single instanton [red], and the density of the quark zero mode [grey]. On the left, there is a single instanton, and both densities fall off as power laws, with $q$ falling off faster. On the right, with the introduction of a magnetic field, the topological charge density is unchanged but the zero mode density is distorted into an asymmetric shape, localized along the direction of the strong magnetic field.
}
\label{fig1}
\end{figure}
In this paper we discuss some features of the spectral problem for fermions in the combined background field of a magnetic field and an instanton. We are motivated by situations in which quarks experience both types of fields, such as in dense astrophysical objects such as neutron stars and magnetars, and in heavy ion collisions such as those at RHIC and at CERN \cite{arXiv:0711.0950,arXiv:0907.1396,arXiv:1111.1949}.  We are also motivated by recent lattice QCD analyses \cite{Buividovich:2009wi,Buividovich:2009my,Abramczyk:2009gb,tom,Braguta:2010ej}, which provide important numerical information about the Dirac spectrum in both QCD and magnetic field backgrounds. Analytically, while the effect of each individual background is very well known, their combined effect turns out to be quite intricate.

\section{General Formalism: Dirac Spectrum}

We briefly review the well-known properties \cite{rubakov} of the Dirac equation for an instanton background {\bf or} a constant magnetic field, since our goal is to discuss what happens when we combine the two background fields. We work in Euclidean four-dimensional spacetime, with the following conventions. We follow the notation of \cite{Jackiw:1977pu}, and express the $4\times 4$ Dirac matrices,
$\gamma_\mu$, for $\mu=1, 2, 3, 4$, in terms of the $2\times 2$ matrices $\alpha_\mu=(\mathbb 1, -i\vec{\sigma})$ and $\bar\alpha_\mu=(\mathbb 1, i\vec{\sigma})=\alpha_\mu^\dagger$, [here $\vec{\sigma}$ are the usual $2\times 2$ Pauli matrices]:
\bea
\gamma_\mu=
\begin{pmatrix}
0&\alpha_\mu\\
\bar\alpha_\mu & 0
\end{pmatrix}
\qquad, \qquad 
\gamma_5=
\begin{pmatrix}
\mathbb 1&0\\
0& -\mathbb 1
\end{pmatrix}
\label{gammas}
\ea
Thus, the Euclidean Dirac operator can be expressed as
\bea
\slashed{\mc D}= \left(\begin{matrix} 0 & \alpha_\mu\mc D_\mu \\\bar\alpha_\mu\mc D_\mu & 0  \end{matrix} \right) \equiv \left(\begin{matrix} 0 & D\\-D^\dagger & 0  \end{matrix} \right) 
\label{dirac}
\ea
where the covariant derivative, $\mc D_\mu=\partial_\mu-i \mc A_\mu$, is written with a hermitean gauge field,  $\mc A_\mu$, and $x_4$ is the Euclidean time coordinate. We write the gauge field $\mc A_\mu$ as a sum of a non-abelian part, $A_\mu$, and an abelian part, $a_\mu$:
\bea
\mc A_\mu=A_\mu+a_\mu
\label{gauge}
\ea
with the respective coupling constants absorbed into the gauge fields.
The Dirac operator is anti-hermitean, so we write (with $\lambda$ real)
\bea
i\slashed{\mc D}\,\psi_\lambda=\lambda \,\psi_\lambda
\label{lambda}
\ea
Since $\{\gamma_5, \slashed{\mc D}\}=0$, we can take $\lambda$ in (\ref{lambda}) to be non-negative, with the negative eigenvalue solutions simply given by $\psi_{-\lambda}=\gamma_5\psi_\lambda$. This means that we can effectively discuss the zero modes ($\lambda=0$) separately, and for the nonzero modes  ($\lambda\neq0$) we consider the squared operator:
\bea
\left(i\slashed{\mc D}\right)^2\psi_\lambda
=
\begin{pmatrix}
DD^\dagger &0\\
0& D^\dagger D
\end{pmatrix}
\psi_\lambda=\lambda^2\psi_\lambda
\label{eig}
\ea
The positive chirality sector, $\chi=+1$, is described by the operator $DD^\dagger$, while the negative chirality sector, $\chi=-1$, is described by the operator $D^\dagger D$. We can write these operators as
\bea
\chi=+1 : \qquad DD^\dagger &=&-\mc D_\mu^2-\mc F_{\mu\nu}\bar\sigma_{\mu\nu}
\label{dd-1}\\
\chi=-1 : \qquad D^\dagger D &=&-\mc D_\mu^2-\mc F_{\mu\nu}\sigma_{\mu\nu}
\label{dd-2}
\ea
We have used $[\mc D_\mu, \mc D_\nu]=-i\mc F_{\mu\nu}$, where $\mc F_{\mu\nu}$ is the field strength associated with the gauge field $\mc A_\mu$, and the spin matrices $\bar\sigma_{\mu\nu}$ and $\sigma_{\mu\nu}$ are defined as
\bea
\bar\sigma_{\mu\nu}&=&\frac{1}{4i}\left(\alpha_\mu\bar\alpha_\nu-\alpha_\nu\bar\alpha_\mu\right)\\
\sigma_{\mu\nu}&=&\frac{1}{4i}\left(\bar\alpha_\mu\alpha_\nu-\bar\alpha_\nu\alpha_\mu\right)
\label{spin}
\ea
In (\ref{dd-1},\ref{dd-2}) we have used the properties \cite{Jackiw:1977pu}: $\bar\alpha_\mu\alpha_\nu=\delta_{\mu\nu}+2 i \sigma_{\mu\nu}$, and $\alpha_\mu\bar\alpha_\nu=\delta_{\mu\nu}+2 i \bar\sigma_{\mu\nu}$.

For non-zero modes [i.e., solutions to (\ref{eig}) with $\lambda\neq 0$], the operators $DD^\dagger$ and $D^\dagger D$ have identical spectra, for any background field. This is simply because we have an invertible map: suppose the 2-component spinor $v$ satisfies $D^\dagger D v=\lambda^2 v$. Then $u= D v$ is clearly an eigenfunction of the other operator, $D  D^\dagger$, with precisely the same eigenvalue: $D  D^\dagger u=D D^\dagger D v =\lambda^2 u$. Similarly, if $u$ satisfies $D D^\dagger u=\lambda^2 u$, then $v=D^\dagger u$ is an eigenstate of $D^\dagger D$ with the same eigenvalue. 
Thus, when $\lambda\neq 0$, we can write the 4-component spinor solution in the form
\bea
\psi_\lambda=
\begin{pmatrix}
u_\lambda \\
-\frac{i}{\lambda}\,D^\dagger u_\lambda
\end{pmatrix}
\qquad {\rm where} \qquad DD^\dagger u_{\lambda}=\lambda^2 u_\lambda
\label{spinor1}
\ea
or in the form
\bea
\psi_\lambda=
\begin{pmatrix}
\frac{i}{\lambda}D v_\lambda\\
v_\lambda
\end{pmatrix}
\qquad {\rm where} \qquad D^\dagger D v_\lambda=\lambda^2 v_\lambda
\label{spinor2}
\ea
This is true for any background field: non-abelian, abelian, or both. 

\subsection{Magnetic field background}\label{magback}

For a constant (abelian) magnetic field, of strength $B$,  pointing in the $x_3$ direction, we have an abelian field strength $f_{12}=B$, and so we find
\bea
\chi=+1 : \qquad DD^\dagger &=&-\mc D_\mu^2-B \sigma_3
\label{dmag1}\\
\chi=-1 : \qquad D^\dagger D &=&-\mc D_\mu^2-B\sigma_3
\label{dmag2}
\ea
where we have used the fact that $\bar\sigma_{12}=\sigma_{12}=\frac{1}{2}\sigma_3$. 
Note that in this case the $2\times 2$ operators $D D^\dagger$ and $D^\dagger D$,  of the two chiral sectors, are the same, and therefore they have identical spectra.
Due to the subtraction term, $-B\sigma_3$, it is possible to have zero modes, and since $D D^\dagger=D^\dagger D$ these zero modes occur in each chiral sector. More explicitly, we can make a Bogomolnyi-style factorization and write 
\bea
-\mc D_\mu^2-B \sigma_3&=&-\partial_3^2-\partial_4^2-\left(\mc D_1\mp i \mc D_2\right)\left(\mc D_1\pm i \mc D_2\right)\pm B -B\sigma_3
\\
&\equiv &-\partial_3^2-\partial_4^2- \mc D_\mp\mc D_\pm \pm B-B\,\sigma_3
\label{bog}
\ea
\begin{figure}[htb]
\includegraphics[scale=0.75]{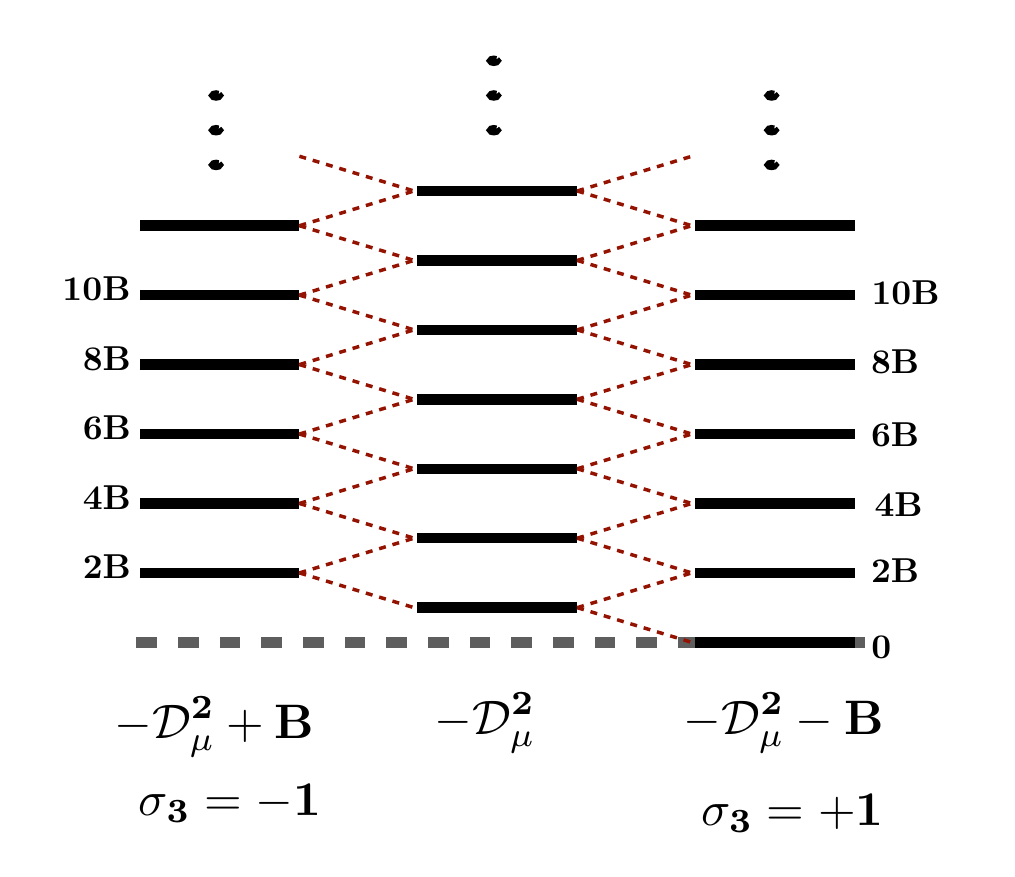}
\caption{A sketch of the form of the Landau level spectrum $\lambda^2$ of the squared Dirac operator $(i\slashed{\mc D})^2$, for a constant abelian magnetic field. The plot is the same for both positive and negative chirality; that is, for the operators $DD^\dagger$ and $D^\dagger D$.  Note that for each chirality the zero mode has spin up (along the magnetic field), and furthermore note that the two-dimensional  operators $\mc D_+\mc D_-$ and $\mc D_-\mc D_+$ are isospectral apart from the zero mode.}
\label{fig2}
\end{figure}
For zero modes, we take $\partial_3=\partial_4=0$, and with $B>0$ we choose the upper signs to ensure normalizable modes. For example, in the symmetric gauge where the abelian gauge field
\bea
a_\mu=\frac{B}{2}(-x_2, x_1, 0, 0)
\label{symmetric}
\ea
 the zero modes can be expressed  in terms of the normalizable solutions to $\left(\mc D_1+ i \mc D_2\right)u=0$:
\bea
\psi_0=g(z_1)e^{-B|z_1|^2/2}
\begin{pmatrix}
1\\0\\0\\0
\end{pmatrix}
\qquad {\rm or} \qquad 
\psi_0=g(z_1)e^{-B|z_1|^2/2}
\begin{pmatrix}
0\\0\\1\\0
\end{pmatrix}
\ea
Here $g(z_1)$ is a holomorphic function of the complex variable $z_1=(x_1+i x_2)/\sqrt{2}$. Both sets of zero modes have spin up, aligned along the $B$ field; this is just the familiar lowest Landau level projection onto spin up states. Note also that the zero modes have the characteristic Gaussian factor in the $(x_1, x_2)$ plane, transverse to the direction of the magnetic field. This factor is the origin of the distortion sketched in the right frame of Fig. 1.

The number of zero modes per unit two-dimensional area [in the $(x_1,x_2)$ plane] is given by the Landau degeneracy factor, the magnetic flux per unit area: $B/(2\pi)$. In fact, even for an inhomogeneous magnetic field $B(x_1, x_2)$, pointing in the $x_3$ direction, the number of zero modes [of each chirality] is determined by the integer part of the magnetic flux (this is the essence of the Aharonov-Casher theorem \cite{aharonov-casher}). For example, on a torus \cite{novikov}:
\bea
N_+=N_-=\frac{1}{2\pi}\int d^2 x\,  B
\label{aharonov}
\ea
The higher Landau level states are the same for both spins, as $(-\mc D_-\mc D_+ +B)$ and $(-\mc D_+\mc D_--B)$ have identical spectra, apart from the lowest level, which only has spin aligned along the magnetic field. The resulting spectrum is sketched in Fig. \ref{fig2}.

\subsection{Instanton background}

For an instanton field, $A_\mu$, the (non-abelian) field strength $F_{\mu\nu}$ is self-dual [that is: $F_{\mu\nu}=\tilde{F}_{\mu\nu}$, where the dual tensor is defined: $\tilde{F}_{\mu\nu}\equiv\frac{1}{2}\epsilon_{\mu\nu\alpha\beta}{F}_{\alpha\beta}$]. Then the anti-self-duality property of $\bar\sigma_{\mu\nu}$ [that is: $\tilde{\bar\sigma}_{\mu\nu}=-\bar\sigma_{\mu\nu}$] implies:
\bea
\chi=+1 : \qquad DD^\dagger &=&-\mc D_\mu^2\\
\chi=-1 : \qquad D^\dagger D &=&-\mc D_\mu^2-F_{\mu\nu}\sigma_{\mu\nu}
\label{ddd-inst}
\ea
Since $-\mc D_\mu^2$ is a positive operator, this means that for an instanton background there can be no zero mode in the positive chirality sector. On the other hand, due to the subtraction term, $-F_{\mu\nu}\sigma_{\mu\nu}$, in $D^\dagger D$, it is possible to have a zero eigenvalue solution in the negative chirality sector, and it has the form
\bea
\psi_0=\begin{pmatrix}
0\\v
\end{pmatrix} \qquad , \quad {\rm where} \qquad Dv=0
\label{zeromode}
\ea
\begin{figure}[htb]
\includegraphics[scale=0.75]{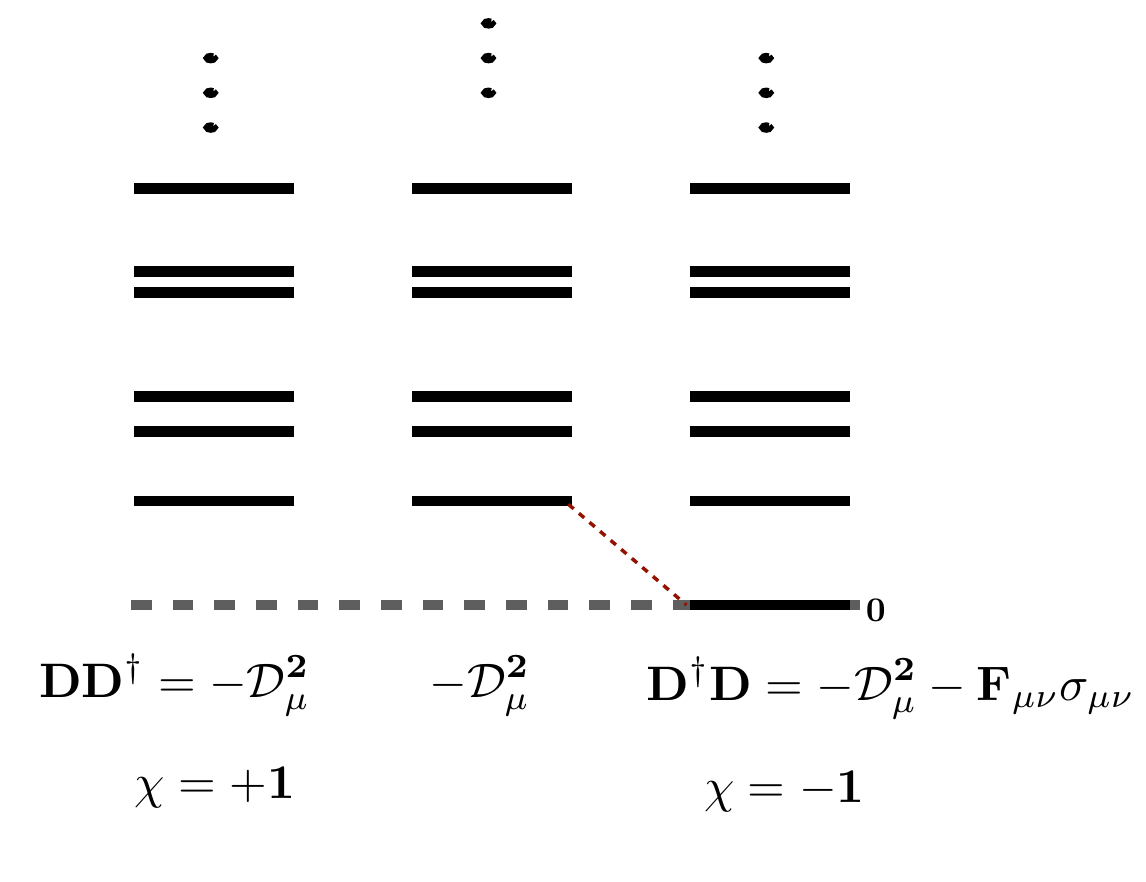}
\caption{A sketch of the form of the spectrum $\lambda^2$ of the squared Dirac operator $(i\slashed{\mc D})^2$, for an instanton field. Note that the operators $DD^\dagger$ and $D^\dagger D$ are isospectral [also with $-{\mathcal D}_\mu^2$] except for a zero mode in the negative chirality sector.}
\label{fig3}
\end{figure}
[For an anti-instanton, an anti-self-dual field with $F_{\mu\nu}=-\tilde{F}_{\mu\nu}$,  the zero mode lies in the positive chirality sector, because $\sigma_{\mu\nu}$ is self-dual: $\tilde{\sigma}_{\mu\nu}=\sigma_{\mu\nu}$.] For a general non-abelian gauge field $A_\mu$, which is neither self-dual nor anti-self-dual, the Atiyah-Singer  index theorem \cite{aps,rubakov} states that the difference between the number of positive and negative chirality zero modes is given by the topological charge of the gauge field:
\bea
N_+-N_-=- \frac{1}{32\pi^2}\int d^4x\,
%\left(
F^a_{\mu\nu}\tilde{F}^a_{\mu\nu}
%\right)
\label{atiyah}
\ea
Here we have written $F_{\mu\nu}=F_{\mu\nu}^a T^a$, with generators normalized as 
${\rm tr}(T^a T^b)=\frac{1}{2}\delta^{ab}$.
For gauge group $SU(2)$, with fermions in the defining representation, we take generators $T^a=\frac{1}{2}\tau^a$ in terms of the Pauli matrices $\vec{\tau}$, and we can  write the single instanton gauge field \cite{Belavin:1975fg}, centered at the origin,  in the regular gauge as 
%[Rubakov's book]
\bea
A_\mu^a=2\frac{ \eta^a_{\mu\nu}\, x_\nu}{x^2+\rho^2}
%A_\mu=-\left(\bar\sigma_\mu\sigma_\nu-\delta_{\mu\nu}\right)\frac{r\, \hat{n}_\nu}{r^2+\rho^2}
\ea
where $\rho$ is the instanton scale parameter, and $\eta^a_{\mu\nu}$ is the self-dual  't Hooft tensor \cite{'tHooft:1976fv,Jackiw:1977pu}. The topological charge density is 
\bea
q(x)=\frac{1}{32\pi^2}\, F^a_{\mu\nu}\tilde{F}^a_{\mu\nu}=\frac{192\, \rho^4}{(x^2+\rho^2)^4}
\label{q}
\ea
There is a single zero mode \cite{'tHooft:1976fv,Schwarz:1977az,Kiskis:1977vh,Brown:1977bj,rubakov}, also localized at the origin, with density:
\bea
|\psi_0|^2=
\frac{64\, \rho^2}{(x^2+\rho^2)^3}
\label{izm}
\ea
These densities both fall off as power laws, with scale set by $\rho$, but the topological charge density is more localized, as indicated in the left-hand frame of Fig. \ref{fig1}. The nonzero modes are given by (\ref{spinor1}) or (\ref{spinor2}), and we note that the spectra are identical in each chiral sector, apart from the zero modes, as is indicated in the sketch in Fig. \ref{fig3}.

\subsection{Combined instanton and magnetic field background}

Physically,  an instanton field projects the zero modes onto a definite chirality, while a constant magnetic field projects the zero modes onto definite spin, aligned along the direction of the magnetic field. When we combine the two background fields, both a non-abelian instanton field $F_{\mu\nu}$ and an abelian magnetic field $f_{12}=B$, there is a competition between the two projection mechanisms, and the outcome depends on their relative magnitude, as we show below. Technically speaking, the instanton zero mode has a specific ansatz form that unifies space-time and color indices, while the magnetic zero modes have a natural holomorphic structure, and these two different ansatz forms do not match one another. The competition between these two ansatz forms makes the combined problem nontrivial. For an instanton field, since the field falls off as a power law, all eigenmodes also fall off with power law behavior. On the other hand, once a constant magnetic field is introduced, for example in the gauge (\ref{symmetric}), all the eigenstates (even those in the higher Landau levels) have a Gaussian factor
$\exp(-B|z_1|^2/2)$ that localizes the modes near the axis of the magnetic field. This is the reason for the distorted density in the right-hand frame of Fig. 1. In the extreme strong magnetic field limit this leads to a dimensional reduction to motion along the magnetic field, with interesting physical consequences such as magnetic catalysis \cite{Gusynin:1995nb} and the chiral magnetic effect \cite{Kharzeev:2004ey,Fukushima:2008xe,Kharzeev:2009fn}.

Concerning zero modes, we begin with a simple but important comment: in the index theorem (\ref{atiyah}), the magnetic field makes no contribution, since with the field strength decomposed into its non-abelian and abelian parts, $\mc F_{\mu\nu}=F_{\mu\nu}+f_{\mu\nu}$, we have
\bea
{\rm tr}\left(\mc F_{\mu\nu}\tilde{\mc F}_{\mu\nu}\right)&=&{\rm tr}\left(F_{\mu\nu}\tilde{F}_{\mu\nu}\right)+({\rm dim})\,f_{\mu\nu}\tilde{f}_{\mu\nu}\\
&=&{\rm tr}\left(F_{\mu\nu}\tilde{F}_{\mu\nu}\right)
\ea
where dim is the dimension of the Lie algebra representation of the non-abelian gauge fields. The cross terms vanish since the Lie algebra generators $T^a$ are traceless, and the $f_{\mu\nu}\tilde{f}_{\mu\nu}$ term vanishes since there is no abelian electric field. For example, if there is no nonabelian field, just an abelian magnetic field, then the topological charge clearly vanishes, and  the index theorem (\ref{atiyah})  is consistent with the fact that $DD^\dagger =D^\dagger D$ for an abelian magnetic background (recall (\ref{dmag1}, \ref{dmag2})), so that there are the same number of zero modes in each chiral sector.
%, and therefore no contribution to the index, which counts the {\it difference}.
Now, with both background fields present, we find
\bea
DD^\dagger &=&-\mc D_\mu^2-B\sigma_3\\
D^\dagger D &=&-\mc D_\mu^2-F_{\mu\nu}\sigma_{\mu\nu}-B\sigma_3
\label{ddd-both}
\ea
Notice that the eigenvalues of $DD^\dagger$ are simply those of the scalar operator $-{\mc D}_\mu^2$, with a spin term $\pm B$, as can be seen clearly in Figure \ref{fig5}.
The fact that there is a subtraction term from the positive operator $-\mc D_\mu^2$ in both chirality sectors tells us that it is possible to have zero modes for each chirality, but their number will depend on the relative magnitude of $F$ and $B$. In the next Section  we study a specific model where we can quantify this precisely. Another important implication is that we may also have some "near-zero-modes", where the $F$ and $B$ subtractions do not exactly cancel the lowest eigenvalue of $-\mc D_\mu^2$, but lower the eigenvalue of $DD^\dagger$ or $D^\dagger D$ to near zero.

\section{Large instanton limit}

In the very strong magnetic field limit, where the magnetic length, $1/\sqrt{B}$, is small compared to the instanton size $\rho$, we expect  a significant distortion of instanton modes and currents. In this limit we can make a simple approximation that reduces the problem to a completely soluble system.

\subsection{Covariantly constant $SU(2)$ instanton and constant abelian magnetic field}

\begin{figure}[htb]
\includegraphics[scale=0.5]{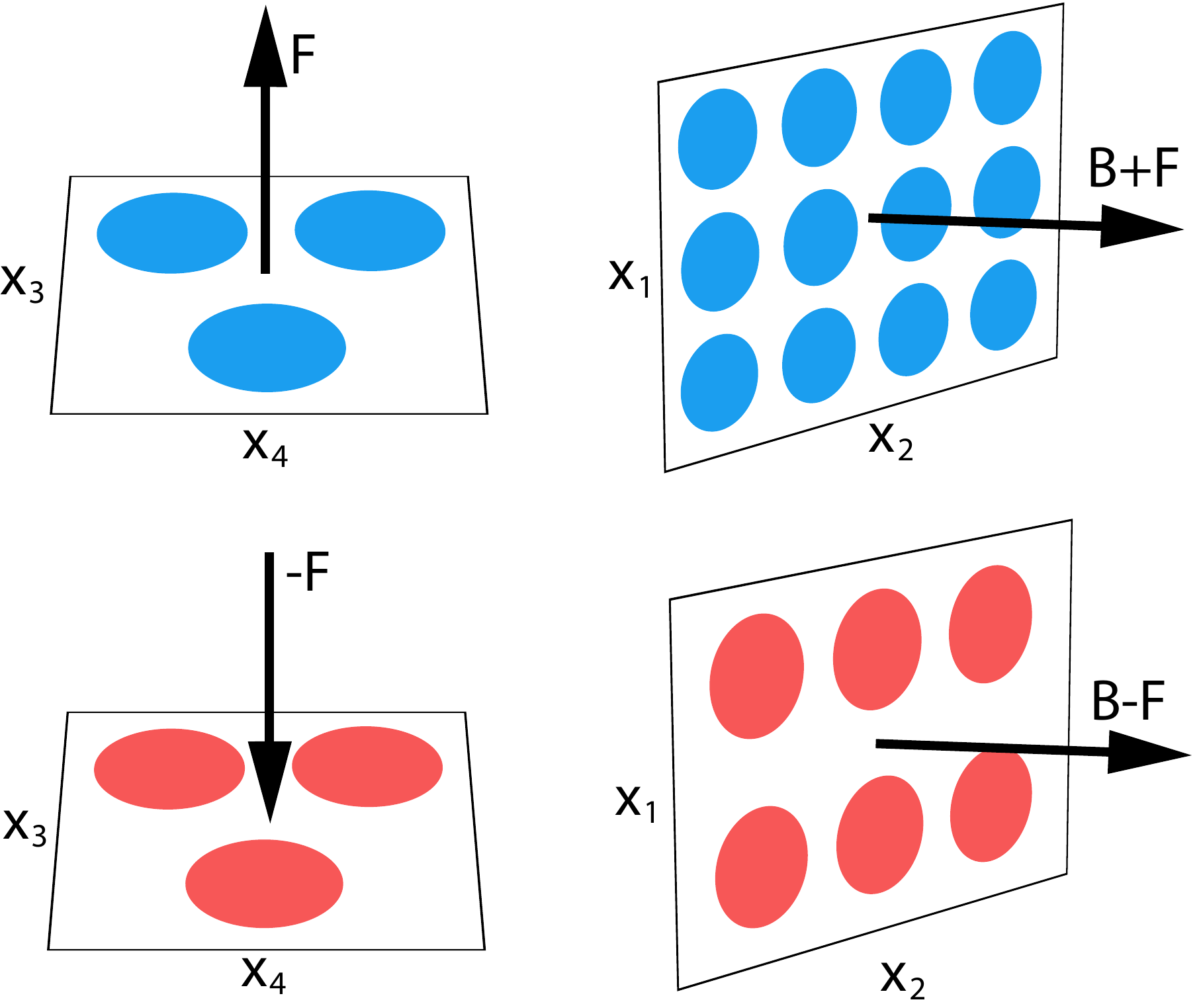}
\caption{Sketch of the effective magnetic field strengths perpendicular to the $(x_1, x_2)$ and $(x_3, x_4)$ planes and the associated zero mode densities which are represented by the colored disks. The colors denote the isospin dependence: blue (upper) stands for $\tau^3=1$, and red (lower) stands for $\tau^3=-1$.  All the zero modes are spin up ($\sigma^3=1$) for B$>$F.  See eqn. (\ref{flux-quanta}) for the case of 4-torus where the magnetic fluxes are quantized and coincide with the number zero modes.}
\label{fig4}
\end{figure}

In the large instanton limit, we expand the instanton gauge field as:
\bea
A_\mu^{a}\approx \frac{2}{\rho^2 } \eta^a_{\mu\nu} x_\nu+\dots
%+O(x^3)
\label{leading}
\ea
To leading order in such a  derivative expansion, the non-abelian gauge configuration $A_\mu^a(x)$ is self-dual and has covariantly constant field strength: $F_{\mu\nu}^a=-\frac{4}{\rho^2} \,\eta_{\mu\nu}^a$. 
In this limit we can make an $SU(2)$ "color" rotation, along with a choice of Lorentz frame,  to make the instanton field diagonal  in the color space (we choose the $\tau^3$ direction), so that the field is self-dual, covariantly constant and quasi-abelian. 
 Defining the instanton scale $F=\frac{2}{\rho^2}$, 
 the combined gauge field, including also the abelian magnetic field as in (\ref{gauge}), can be written as: 
\bea
\mc A_\mu=-\frac{F}{2}(-x_2, x_1, -x_4, x_3)\tau^3+\frac{B}{2}(-x_2, x_1, 0, 0)\mathbb{1}_{2\times2}
\label{gauge2}
\ea
This gauge field is fully diagonal and moreover is linear in $x_\mu$, so the problem is analytically soluble (this is the basic premise of the derivative expansion).
The only nonzero entries of the field strength tensor are
\bea
\mc F_{12}&=&-F\tau^3+B\mathbb{1}=
\begin{pmatrix}
B-F &0\cr
0&B+F
\end{pmatrix}
\nonumber\\
\mc F_{34}&=&-F\tau^3
=
\begin{pmatrix}
-F &0\cr
0&+F
\end{pmatrix}
\ea
In the absence of the magnetic field the field strength is self-dual, $\mc F_{12}=\mc F_{34}$, but a nonzero magnetic field breaks this symmetry. The topological charge density is (recall the normalization of the generators)
\bea
\frac{1}{32\pi^2} \,\mc F_{\mu\nu}^a \tilde{\mc F}_{\mu\nu}^a=\frac{4(2F)^2}{32\pi^2}=\frac{F^2}{2\pi^2}
\label{tc}
\ea
To study the Dirac spectrum we consider the $2\times 2$ operators $DD^\dagger$ and $D^\dagger D$  in (\ref{dd-1},\ref{dd-2}). Notice first that
\bea
\mc F_{\mu\nu}\,\bar\sigma_{\mu\nu}&=&\left(\mc F_{12}-\mc F_{34}\right)\sigma_3
%=B\sigma_3
\\
\mc F_{\mu\nu}\,\sigma_{\mu\nu}&=&\left(\mc F_{12}+\mc F_{34}\right)\sigma_3
\ea
It is convenient to factor the 4-dimensional Euclidean space and consider separately the $(x_1, x_2)$ plane and the $(x_3, x_4)$  plane, as sketched in Figure \ref{fig4}. Then in the $(x_1, x_2)$ plane we have a (relativistic) Landau level problem with effective field strength $(B-F)$ in the $\tau^3=+1$ sector, and with effective field strength $(B+F)$ in the $\tau^3=-1$ sector. In the $(x_3, x_4)$ plane we also have a (relativistic) Landau level problem, now with effective field strength $-F$ in the $\tau^3=+1$ sector, and with effective field strength $F$ in the $\tau^3=-1$ sector. In the $(x_1, x_2)$ plane the sign of the effective field strength depends on which of $B$ or $F$ is larger, so we consider separately the cases $B>F$ or $B<F$. 

\subsubsection{Strong magnetic field limit: $B>F$}

When $B>F$, both $(B-F)$ and $(B+F)$ are positive. Thus, each color component of $\mc F_{12}$ is associated with a positive "magnetic" field. On the other hand, for $\mc F_{34}$, the $\tau^3=+1$ sector has  a negative field strength, while the  $\tau^3=-1$ sector has  a  positive  field strength. 

We first consider the $\tau^3=+1$ case. Then $\mc F_{12}=(B-F)$, $\mc F_{34}=-F$,
$\mc F_{\mu\nu}\bar\sigma_{\mu\nu}=B\sigma_3$, and $\mc F_{\mu\nu}\sigma_{\mu\nu}=(B-2F)\sigma_3$. With  a positive field strength the normalizable zero state is given by $\left(\mc D_1+i\mc D_2\right)u=0$.
But since  $\mc F_{34}$ is negative, we factorize the corresponding covariant derivatives in the opposite order, in order to obtain a normalizable state annihilated by $\left(\mc D_3-i\mc D_4\right)$. Thus, we have:
\bea
\chi=+1:\qquad D D^\dagger&=& -\left(\mc D_1-i\mc D_2\right)\left(\mc D_1+i\mc D_2\right) +\mc F_{12} 
- \left(\mc D_3+i\mc D_4\right)\left(\mc D_3-i\mc D_4\right)-\mc F_{34}-B \sigma_3\nonumber\\
&=& -\left(\mc D_1-i\mc D_2\right)\left(\mc D_1+i\mc D_2\right) 
- \left(\mc D_3+i\mc D_4\right)\left(\mc D_3-i\mc D_4\right)
+B-B \sigma_3
\label{bf++}\\
\chi=-1:\qquad  D^\dagger D&=& -\left(\mc D_1-i\mc D_2\right)\left(\mc D_1+i\mc D_2\right) +\mc F_{12} 
- \left(\mc D_3+i\mc D_4\right)\left(\mc D_3-i\mc D_4\right)-\mc F_{34}-(B-2F) \sigma_3\nonumber\\
&=& -\left(\mc D_1-i\mc D_2\right)\left(\mc D_1+i\mc D_2\right) 
- \left(\mc D_3+i\mc D_4\right)\left(\mc D_3-i\mc D_4\right)
+B-(B-2F) \sigma_3
\label{bf+-}
\ea
This shows that there is a zero mode,  when the spin term $B \sigma_3$ cancels the $B$ term from the Bogomolnyi factorization of the covariant derivative term. This occurs  in the positive chirality sector, $\chi=+1$, and  with spin up: $\sigma_3=+1$.
\begin{figure}[htb]
\includegraphics[scale=0.75]{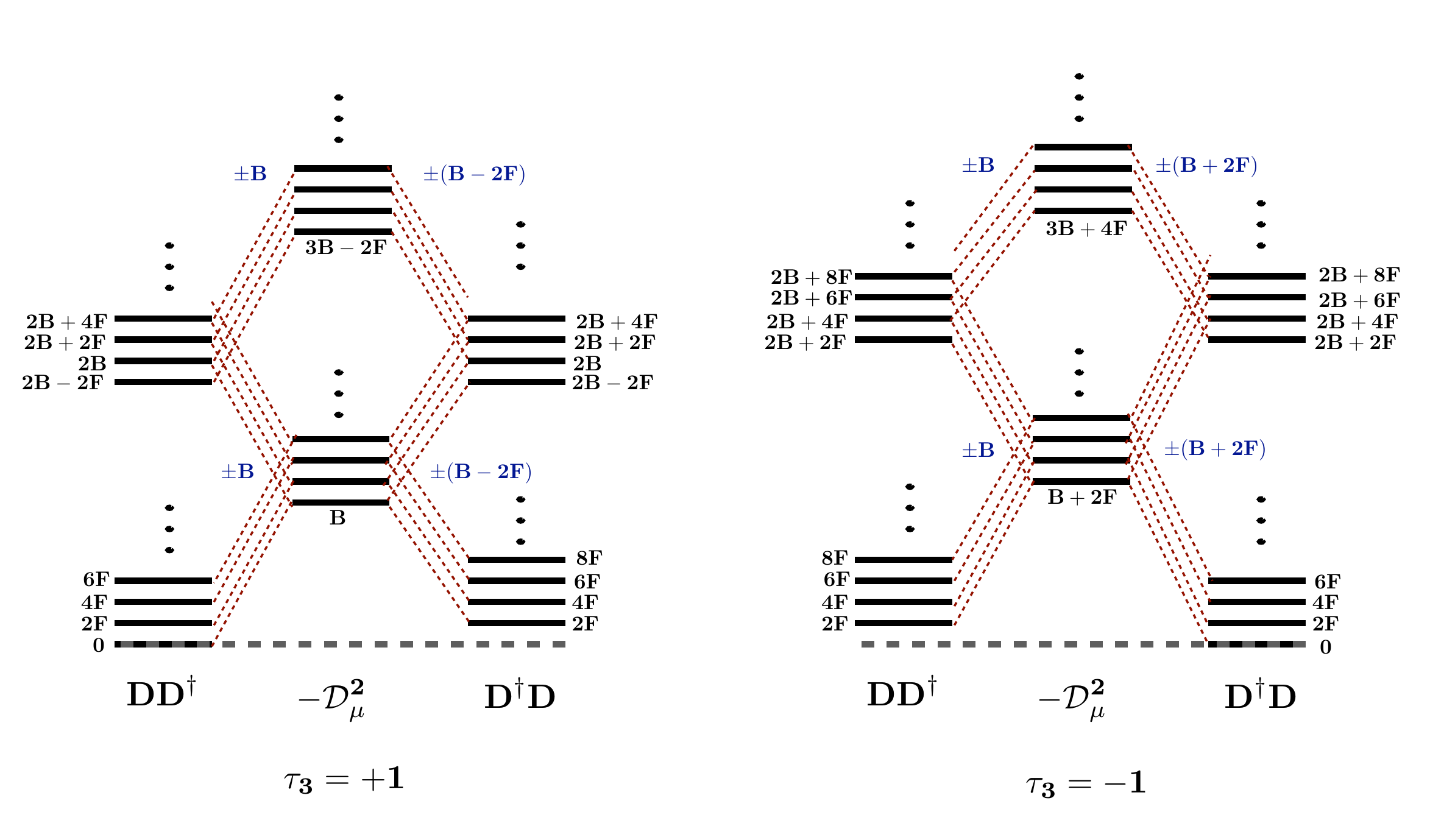}
\caption{Spectrum of the squared Dirac operator $(i\slashed{\mc D})^2$, for both a strong magnetic field and instanton background, with $B\gg F$, as derived from equations (\ref{bf++}, \ref{bf+-}, \ref{bf-+}, \ref{bf--}). Note that for each "color", $\tau_3=\pm 1$, the operators $DD^\dagger$ and $D^\dagger D$ are isospectral except for a zero mode. For $\tau_3=+1$, this zero mode is in the positive chirality sector, while for $\tau_3=-1$, this zero mode is in the negative chirality sector. The spin-projection steps relating the spectra of $DD^\dagger$,  $D^\dagger D$ and ${\mathcal D}_\mu^2$, steps of $\pm B$, $\pm(B-2F)$ and $\pm(B+2F)$, are indicated in the figures.}
\label{fig5}
\end{figure}

Now consider the $\tau^3=-1$ case. 
Then $\mc F_{12}=(B+F)$, $\mc F_{34}=F$,
$\mc F_{\mu\nu}\bar\sigma_{\mu\nu}=B\sigma_3$, and $\mc F_{\mu\nu}\sigma_{\mu\nu}=(B+2F)\sigma_3$.
All field strengths are positive, so we  write
\bea
\chi=+1:\qquad D D^\dagger&=& -\left(\mc D_1-i\mc D_2\right)\left(\mc D_1+i\mc D_2\right) +\mc F_{12} 
- \left(\mc D_3-i\mc D_4\right)\left(\mc D_3+i\mc D_4\right)+\mc F_{34}-B \sigma_3\nonumber\\
&=& -\left(\mc D_1-i\mc D_2\right)\left(\mc D_1+i\mc D_2\right) 
- \left(\mc D_3-i\mc D_4\right)\left(\mc D_3+i\mc D_4\right)
+(B+2F)-B \sigma_3
\label{bf-+}\\
\chi=-1:\qquad  D^\dagger D&=& -\left(\mc D_1-i\mc D_2\right)\left(\mc D_1+i\mc D_2\right) +\mc F_{12} 
- \left(\mc D_3-i\mc D_4\right)\left(\mc D_3+i\mc D_4\right)+\mc F_{34}-(B+2F) \sigma_3\nonumber\\
&=& -\left(\mc D_1-i\mc D_2\right)\left(\mc D_1+i\mc D_2\right) 
- \left(\mc D_3-i\mc D_4\right)\left(\mc D_3+i\mc D_4\right)
+(B+2F)-(B+2F) \sigma_3
\label{bf--}
\ea
This shows that there is a zero mode, but now in the opposite chirality sector, $\chi=-1$, and also with spin up: $\sigma_3=+1$.

To summarize: when $B>F$, the $\tau_3=+1$ color sector has  spin up zero modes with positive chirality, while the $\tau_3=-1$ color sector has  spin up zero modes with negative chirality. 
We can count the number of zero modes in each chirality sector by simply taking the product of the Landau degeneracy factors for the $(x_1, x_2)$ and $(x_3, x_4)$ planes, with the corresponding effective magnetic field strengths. Therefore, 
%The $\tau_3=+1$ color sector has  spin up zero modes with negative chirality, while the $\tau_3=-1$ color sector has  spin up zero modes with positive chirality. 
the corresponding Landau degeneracy factors give the zero-mode number densities (i.e, the number per unit volume):
\bea
\chi=+1:\qquad n_+&=&\frac{(B-F)}{2\pi}\,\frac{F}{2\pi} \qquad (\tau_3=+1\quad, \quad \sigma_3=+1)
\label{bf-landau-1}\\
\chi=-1:\qquad n_-&=&\frac{(B+F)}{2\pi}\,\frac{F}{2\pi}  \qquad (\tau_3=-1\quad, \quad \sigma_3=+1)
\label{bf-landau-2}
\ea
The index (density) is the difference, 
\bea
n_+-n_-=-\frac{F^2}{2\pi^2}
\ea
in agreement with the general index theorem (\ref{atiyah}), in view of (\ref{tc}). 
We also note that the total number density of zero modes
\bea
n_++n_-=\frac{BF}{2\pi^2}
\ea
is linearly proportional to the magnetic field strength $B$. This is in agreement with numerical lattice gauge theory results \cite{tom}.

\subsubsection{Weak magnetic field limit: $B<F$}

Even though this limit is outside the derivative expansion regime that originally motivated the quasi-abelian covariantly constant gauge field ansatz (\ref{leading}), it is still instructive to see how the Dirac spectrum changes for such a field when $B<F$. The difference is that now the $\tau^3=+1$ effective "magnetic" field $(B-F)$ is negative, so the factorization should be done in the opposite order also for the $(x_1, x_2)$ plane.

Consider the $\tau^3=+1$ case first. 
Then $\mc F_{12}=(B-F)$, $\mc F_{34}=-F$,
$\mc F_{\mu\nu}\bar\sigma_{\mu\nu}=B\sigma_3$, and $\mc F_{\mu\nu}\sigma_{\mu\nu}=(B-2F)\sigma_3$. Since  $\mc F_{12}$ has the opposite sign from before, we factorize the corresponding covariant derivatives in the opposite order, in order to obtain a normalizable state annihilated by $\left(\mc D_1-i\mc D_2\right)$. Thus, we have:
\bea
\chi=+1:\qquad D D^\dagger&=& -\left(\mc D_1+i\mc D_2\right)\left(\mc D_1-i\mc D_2\right) -\mc F_{12} 
- \left(\mc D_3+i\mc D_4\right)\left(\mc D_3-i\mc D_4\right)-\mc F_{34}-B \sigma_3\nonumber\\
&=& -\left(\mc D_1+i\mc D_2\right)\left(\mc D_1-i\mc D_2\right) 
- \left(\mc D_3+i\mc D_4\right)\left(\mc D_3-i\mc D_4\right)
-(B-2F)-B\sigma_3\\
\chi=-1:\qquad  D^\dagger D&=& -\left(\mc D_1+i\mc D_2\right)\left(\mc D_1-i\mc D_2\right) -\mc F_{12} 
- \left(\mc D_3+i\mc D_4\right)\left(\mc D_3-i\mc D_4\right)-\mc F_{34}-(B-2F) \sigma_3\nonumber\\
&=& -\left(\mc D_1+i\mc D_2\right)\left(\mc D_1-i\mc D_2\right) 
- \left(\mc D_3+i\mc D_4\right)\left(\mc D_3-i\mc D_4\right)
-(B-2F)-(B-2F) \sigma_3
\ea
This shows that there is a zero mode,  in the negative chirality sector, $\chi=-1$, and with spin down: $\sigma_3=-1$.

Now consider the $\tau^3=-1$ case. 
This is exactly as before, with $\mc F_{12}=(B+F)$, $\mc F_{34}=F$,
$\mc F_{\mu\nu}\bar\sigma_{\mu\nu}=B\sigma_3$, and $\mc F_{\mu\nu}\sigma_{\mu\nu}=(B+2F)\sigma_3$.
Thus, we can write $DD^\dagger$ and $D^\dagger D$ exactly as in (\ref{bf-+}, \ref{bf--}), and we see that, as 
before, the zero modes are in the negative chirality sector, with spin up.

To summarize: when $B<F$, the $\tau_3=+1$ color sector has  spin up zero modes with negative chirality, while the $\tau_3=-1$ color sector has  spin down zero modes also with negative chirality. 
%The $\tau_3=+1$ color sector has  spin up zero modes with negative chirality, while the $\tau_3=-1$ color sector has  spin up zero modes with positive chirality. 
Counting the corresponding Landau degeneracy factors we obtain:
\bea
\chi=+1:\qquad n_+&=&0
\label{fb-landau-1}\\
\chi=-1:\qquad n_-&=&\begin{cases}
\frac{(B+F)}{2\pi}\,\frac{F}{2\pi}\quad , \quad (\tau_3=-1\quad, \quad \sigma_3=+1)\cr
\frac{(-B+F)}{2\pi}\,\frac{F}{2\pi}\quad , \quad (\tau_3=+1\quad, \quad \sigma_3=-1)
\end{cases}
\label{fb-landau-2}
\ea
The total number density of negative chirality zero modes is (note that the $B$ dependence cancels)
\bea
n_-=\frac{F^2}{2\pi^2}
\ea
As before, the index (density) is given by  the difference, 
\bea
n_+-n_-=-\frac{F^2}{2\pi^2}
\ea
in agreement with the general index theorem (\ref{atiyah}). 
In this case the total number density of zero modes
\bea
n_++n_-=\frac{F^2}{2\pi^2}
\ea
which is independent of the magnetic field strength $B$, and equal to (minus) the index. 

\subsection{Physical picture}

These results lead to the following simple physical picture. The instanton tries to generate a chirality imbalance but is neutral to the spin, whereas the magnetic field tries to generate a spin imbalance but does not affect the chirality. Depending on which is stronger, the zero modes have either a definite spin with a chirality imbalance ($B>F$), or a definite chirality with a spin imbalance ($F>B$). 
Also we see that in the former case, the total number of zero modes scales with $B$ and is not equal to the index, while in the latter case it is independent of the $B$ field, and is equal to the magnitude of the index (even though the field is not self-dual).

More explicitly, for the $B>F$ case, consider starting with just a strong magnetic field $B$, later  turning on a weak instanton field.  Without the instanton field, the zero modes and their degeneracy are given by the Aharonov-Casher theorem (\ref{aharonov}), so that the zero mode density is the Landau degeneracy factor $B/(2\pi)$ for each chirality sector. 
All the zero modes are spin up, as is familiar for the lowest Landau level (see Fig. \ref{fig2}). 
There is an equal number of positive and negative chirality zero-modes, 
 which is consistent with the index theorem, since the topological charge vanishes for a constant $B$ field. 
Now consider turning on an instanton field $F$, with $B>F>0$. We see from (\ref{bf-landau-1},\ref{bf-landau-2}) that the effect of the instanton is to flip some of the chiralities: $\left(\frac{F}{2\pi}\right)^2$ positive chirality modes become negative chirality modes,  leading to a chirality imbalance of $\frac{F^2}{2\pi^2}$, in agreement with the index theorem (\ref{atiyah}). On the other hand, the total number of zero modes, $\frac{BF}{2\pi^2}$, grows linearly with the magnetic field when $F$ is nonzero.

On the other hand, when $F>B$ we have the following physical picture. With just the instanton, there are $\frac{F^2}{2\pi^2}$ zero modes, all with negative chirality. Of these,  $\left(\frac{F}{2\pi}\right)^2$ have spin up, and $\left(\frac{F}{2\pi}\right)^2$ have spin down. Now turn on a  magnetic field $B$, with $0<B<F$. The effect of the magnetic field is to flip some of the down spins to up spins, without affecting the chirality. From (\ref{fb-landau-1},\ref{fb-landau-2}) we see that $\frac{B\,F}{(2\pi)^2}$ zero modes have their spin flipped, leading to a spin imbalance, without creating a chirality imbalance. Thus, the index is still equal (in magnitude) to the total number of zero modes.

\subsection{Creation and Annihilation Operator Formalism on the Four-Torus}

To count degeneracies it is convenient to introduce torus boundary conditions, $x_\mu \sim x_\mu+L_\mu$, and so to study this configuration on a 4-torus. The nontrivial homotopy group of the torus leads to a well defined topological structure for a constant instanton field \cite{'tHooft:1981sz, vanBaal:1982ag, vanBaal:1984ar,vanBaal:1995eh}. 
Furthermore, it allows a direct comparison with some recent lattice results \cite{Buividovich:2009wi,Abramczyk:2009gb,tom}, in particular the counting of zero modes. 
We use torus boundary conditions, namely that the gauge field (\ref{gauge2}) is periodic up to a gauge transformation:
\bea
A_\mu(x_\nu+L_\nu)=\Omega^{-1}_\nu(x)(A_\mu(x_\nu)-i \partial_\mu) \Omega_\nu(x)
\ea 
where $\nu$ denotes the shifted coordinate and $\Omega_\nu$ is the associated cocyle. A general treatment of  non-abelian gauge fields on the torus can be found in \cite{'tHooft:1981sz, vanBaal:1982ag, vanBaal:1984ar,vanBaal:1995eh}. 

As in the standard analysis of the constant magnetic field problem, it is useful to work with complex coordinates. The 4-torus can be imagined as two orthogonal 2d planes, with appropriate periodicity conditions,  parameterized by two complex coordinates:
\bea
z_1&\equiv&\frac{x_1+ix_2}{\sqrt{2}}\qquad , \qquad z_2\equiv\frac{x_3+ix_4}{\sqrt{2}}
\label{zeds}
\ea
In these coordinates, the $2\times 2$ Dirac operator $D\equiv\alpha_\mu \mc D_\mu$, defined in (\ref{dirac}), is:
\bea
D=-i\sqrt{2}\left(\begin{matrix} \bar\del_2-\frac{F}{2}  z_2 && \del_1-\frac{B-F}{2} \bar z_1 \\ \\ 
\bar\del_1+\frac{B-F}{2} z_1 && -\del_2-\frac{F}{2} \bar z_2\end{matrix} \right)\otimes \mathbb I_+
-i\sqrt{2}\left(\begin{matrix}  \bar\del_2+\frac{F}{2}  z_2 &&\del_1-\frac{B+F}{2} \bar z_1\\ \\
 \bar\del_1+\frac{B+F}{2} z_1 && -\del_2+\frac{F}{2} \bar z_2\end{matrix} \right)\otimes \mathbb I_-
 \label{d1}
\ea
Here $\mathbb I_\pm$ denote the color projection matrices. 
Following the standard treatment for a constant magnetic field, we define the ladder operators: 
\bea
a_1=-i\sqrt{2}\left(\bar\del_1+\frac{B-F}{2} z_1\right) \qquad , \qquad \text{for $\mathbb I_+$} \nn
\ta_1=-i\sqrt{2}\left(\bar\del_1+\frac{B+F}{2} z_1\right) \qquad ,\qquad \text{for $\mathbb I_-$}\nn
a_2=-i\sqrt{2}\left(\del_2+\frac{F}{2} \bar z_2\right)\qquad ,\qquad \text{for $\mathbb I_+$} \nn
\ta_2=-i\sqrt{2}\left(\bar\del_2+\frac{F}{2}  z_2\right)  \qquad ,\qquad \text{for $\mathbb I_-$}
\label{as}
\ea
These satisfy the commutation relations
\bea&[a_1,a_1^\dagger]=2(B-F)& \nn
&[\ta_1,\ta_1^\dagger]=2(B+F)&\nn
&[a_2,a_2^\dagger]=2F&\nn
&[\ta_2,\ta_2^\dagger]=2F&
\label{commutation}
\ea
It is now explicit that, for each color, we have a set of two independent Landau level problems and the Landau levels are governed independently by the annihilation-creation operators $(a_i, a^\dagger_i)$ for $\tau_3=+1$, and 
$(\ta_i,\ta^\dagger_i)$ for $\tau_3=-1$. We denote the associated number operators as:
\bea
&a_1^\dagger a_1=2(B-F)\,N_1 \qquad , \qquad a_2^\dagger a_2=2F\,N_2 &\nn
&\ta_1^\dagger \ta_1=2(B+F)\,\tilde N_1 \qquad , \qquad \ta_2^\dagger \ta_2=2F\,\tilde N_2&
\ea
One should keep in mind that within our large instanton approximation, in which $B\gg F$, the Landau levels for the fields perpendicular to  the $(x_1,x_2)$ plane have  greater degeneracy than the ones for the fields perpendicular to  the $(x_3,x_4)$ plane. Expressed in terms of these ladder operators, the Dirac operator $D\equiv\alpha_\mu \mc D_\mu$, defined in (\ref{dirac}), is given by:
\bea
D=\left(\begin{matrix} a_2^\dagger & a_1^\dagger \\  a_1 & -a_2  \end{matrix} \right)\otimes \mathbb I_++\left(\begin{matrix} \ta_2 & \ta_1^\dagger \\  \ta_1 & -\ta_2^\dagger  \end{matrix} \right)\otimes \mathbb I_-
\label{d2}
\ea
To find the Dirac eigenvalues we consider $(i\slashed{\mc D})^2$, as in (\ref{eig}), which means we require:
\bea
%-\slashed{\mc D}^2&=&\left(\begin{matrix} DD^\dagger & 0 \\  0 & D^\dagger D  \end{matrix} \right)\nn
DD^\dagger&=&\left( 2(B-F)N_1+2FN_2+\left(\begin{matrix} 0 & 0 \\  0 & 2B  \end{matrix} \right) \right)\otimes \mathbb I_++\left(2(B+F)\tilde N_1+2F\tilde N_2+\left(\begin{matrix} 2F & 0 \\  0 & 2(B+F)  \end{matrix} \right)\right)\otimes \mathbb I_-\nn\nn
D^\dagger D&=&\left((2(B-F)N_1+2FN_2+\left(\begin{matrix} 2F & 0 \\  0 & 2(B-F)  \end{matrix} \right)\right)\otimes \mathbb I_++\left(2(B-F)\tilde N_1+2F\tilde N_2+\left(\begin{matrix} 0 & 0 \\  0 & 2(B+2F)  \end{matrix} \right)\right)\otimes \mathbb I_-\nn
\label{ddd}
\ea
Thus, $(i\slashed{\mc D})^2$ is fully diagonalized and we have a complete description of the entire spectrum in terms of elementary harmonic oscillator number operators.
 Notice that, in addition to the Landau level operators (\ref{as}), we can define another set of ladder operators which commute with all the $a_i$ and $\ta_i$, and therefore with the Dirac operator $\slashed{\mc D}$:
\bea
b_1=-i\sqrt{2}\left(\bar\del_1-\frac{B-F}{2} z_1\right) \qquad , \qquad \text{for $\mathbb I_+$} \nn
\tb_1=-i\sqrt{2}\left(\bar\del_1-\frac{B+F}{2} z_1\right) \qquad ,\qquad \text{for $\mathbb I_-$}\nn
b_2=-i\sqrt{2}\left(\del_2-\frac{F}{2} \bar z_2\right)\qquad ,\qquad \text{for $\mathbb I_+$} \nn
\tb_2=-i\sqrt{2}\left(\bar\del_2-\frac{F}{2}  z_2\right)  \qquad ,\qquad \text{for $\mathbb I_-$}
\label{bs}
\ea
These operators satisfy the same commutation relation as the Landau level ladder operators:
\bea&[b_1,b_1^\dagger]=2(B-F)& \nn
&[\tb_1,\tb_1^\dagger]=2(B+F)&\nn
&[b_2,b_2^\dagger]=2F&\nn
&[\tb_2,\tb_2^\dagger]=2F&
\ea
These operators generate magnetic translations and so characterize the degeneracies of the Landau levels. With torus boundary conditions, the degeneracy is finite and given by the net flux quanta through the period parallelogram of the corresponding 2-torus. In other words, \textit {each} Landau level is a finite dimensional representation of the magnetic translation group \cite{Zak:1964zz,novikov,AlHashimi:2008hr} with the dimension being equal to the flux. For simplicity of notation we take all four periods to be equal, denoted  by $L$. Then for each of the two 2-tori [and for each of the two colors] we have the flux quantization conditions:  
\bea
&(B-F)\,L^2=2\pi(N-M)& \qquad , \qquad\text{for $\mathbb I_+$} \nn
&(B+F)\,L^2=2\pi(N+M)& \qquad , \qquad \text{for $\mathbb I_-$} \nn
&F\,L^2=2\pi M& \qquad , \qquad \text{for $\mathbb I_+$} \nn
&F\,L^2=2\pi M & \qquad , \qquad \text{for $\mathbb I_-$} 
\label{flux-quanta}
\ea
Here $N$ and $M$ are nonzero integers and each Landau level has  degeneracy factor 
\bea
N_\pm = M \left| N \mp M\right| \qquad , \qquad {\rm for}\quad \tau_3=\pm 1
\label{deg}
\ea
since we simply multiply the degeneracy factors for the two independent 2-tori for the $(x_3,x_4)$  and  $(x_1,x_2)$ planes. 
The explicit states can be generated by acting on the ground state with the various creation operators, and the associated wavefunctions can be written in terms of elliptic functions, as is familiar \cite{novikov,AlHashimi:2008hr}.  We will concentrate instead on  the full eigenvalue spectrum of $(i\slashed{\mc D})^2$.

\subsection{Zero modes}

As described above, we treat separately the cases where $B>F$ and $B<F$.
The torus boundary conditions allow us to express the degeneracies exactly since the relevant fluxes are integers. 

\subsubsection{$B>F$} 

As explained above, with just the magnetic field present, there are only spin-up zero modes with both chiralities. Let us denote the degeneracy as $2N$, where $ B L^2/2 \pi=N$. The factor 2 is due to color. 
Now consider turning on an instanton field $F$, so that $B>F>0$, with $F=2\pi M/L^2$ as in (\ref{flux-quanta}).
The zero modes can be easily constructed by looking at (\ref{ddd}), with $N_1=N_2=\tilde N_1=\tilde N_2=0$. It is clear from (\ref{ddd})  that there only two sectors where one can have zero modes. Their spin and chirality can be read off directly. Furthermore, their degeneracies are fixed by (\ref{flux-quanta}). As a result, we see that there are $(N-M) M$ spin up, positive chirality zero modes for the $\mathbb I_+$ color sector, and  $(N+M) M$ spin up, negative chirality zero modes for the $\mathbb I_-$ color sector. Here $M$ is the instanton flux and $N$ is the magnetic flux, from (\ref{flux-quanta}). The index of the Dirac operator is given by the {\it difference}:
\bea
{\rm index}(\slashed{\mc D})\equiv N_+-N_- &=& (N-M)M-(N+M)M\nn
&=&-2M^2\nn
&=&-\frac{F^2\, L^4}{2\pi^2}
%&=&-\frac{1}{8 \pi^2}\int \mc F_{\mu\nu} \tilde{\mc F}^{\mu\nu}
\ea
in agreement with (\ref{atiyah}) and (\ref{tc}).
The effect of the instanton is to flip the chirality of $M^2$ fermion zero-modes, resulting in a chirality difference of $2M^2$.
Also, the total number of zero modes is:
\bea
\text{total number of zero modes}&=&(N+M)M+(N-M)M\nn
&=&2NM\nn
&=&\frac{B\, F\, L^4}{2\pi^2} 
%\propto B\,F
\ea
in agreement with (\ref{bf-landau-1},\ref{bf-landau-2}). We see that the total number of zero modes is linearly proportional to the magnetic field strength. This agrees with the lattice results \cite{tom}. The functional forms of these zero modes can again be constructed in the same manner as above.

\subsubsection{$F>B$}

It is a straightforward exercise to construct the analogue of (\ref{ddd}) for $F> B$, from which  the zero modes can easily be constructed.  
First, consider just the instanton field, so $F$ is nonzero, but $B=0$.  In this case, there are $2M^2$ zero modes. All of them are positive chirality, in agreement with the vanishing theorem \cite{Brown:1977bj} that states that for a self-dual field the index is equal to the total number of zero modes, since all zero-modes have the same chirality. Also, as expected, there is no preference in spin and there are equal number of spin up and down zero modes.

When we turn on a weak magnetic field, so that $B=2\pi N/L^2$ is nonzero, the effect of the magnetic field is to flip some of the down spins to up spins. Now there are $(N+M)M$ spin up zero modes, and  $(M-N)M$ spin down zero modes. 
The zero modes are still positive chirality, since the magnetic field does not flip chirality. Therefore the index is still equal to the total number of zero modes even though the gauge field is not self-dual anymore with the magnetic field. 
\bea
{\rm index}(\slashed{\mc D})\equiv N_+-N_- &=& 0-2M^2\nn
&=&-2M^2\nn
&=&-\frac{F^2\, L^4}{2\pi^2}
%\nn
%&=&\frac{1}{8 \pi^2}\int F_{\mu\nu} \tilde F^{\mu\nu}
\ea
In this case, the \textit{total} number of zero modes is also equal to the index:
\bea
\text{total number of zero modes}&=&(M-N)M+(M+N)M\nn
&=&2M^2
%&=&\frac{B\, F\, L^4}{2\pi^2} 
%\propto B\,F
\ea
As a result we explicitly get the same simple physical picture as described above. Depending on whether the magnetic field or the instanton is stronger, the zero modes have either a definite spin with a chirality imbalance ($B>F$) or a definite chirality  with a spin imbalance ($F>B$).  
The advantage of the torus construction is two-fold: first, the fluxes are integers and the wavefunctions have simple expressions in terms of elliptic functions; and second, the counting of zero modes agrees with the numerical lattice QCD results.

\subsection{Landau levels}
From the quantum mechanical supersymmetry relations (\ref{spinor1}, \ref{spinor2}) we can construct the full spinor solution from $\ket{\psi_R}$, a two-component spinor satisfying:
\bea
&DD^\dagger\ket{\psi_R}=\lambda^2\ket{\psi_R}&\nn
%&\ket{\psi_R}=\left( \begin{matrix} \ket{n_1,n_2}\\ \ket{n_1-1,n_2-1}\end{matrix}\right)&\nn
%&\lambda_{n_1,n_2}=\sqrt{2(B-F)\,n_1+2F\,n_2}&
\ea
Since $DD^\dagger$ commutes with $\sigma_3$, from (\ref{ddd-both}), we can choose $\ket{\psi_R}$ to be either spin up or spin down.
In the $\tau_3=+1$ sector, we find
\bea
%&DD^\dagger\ket{\psi_R}=\lambda^2\ket{\psi_R}&\nn
\ket{\psi_{R\uparrow}}&=&\left( \begin{matrix} \ket{n_1,n_2}\\ 0\end{matrix}\right)\qquad, \qquad
\lambda_{n_1,n_2}=\sqrt{2(B-F)\,n_1+2F\,n_2} \nonumber\\
\ket{\psi_{R\downarrow}}&=&\left( \begin{matrix} 0 \\ \ket{n_1-1,n_2-1}\end{matrix}\right)\qquad, \qquad
\lambda_{n_1,n_2}=\sqrt{2(B-F)\,n_1+2F\,n_2}
\ea
For the spin up states, $n_1,n_2\in\{0, 1,2,3...\}$, excluding the case where both $n_1=n_2=0$, while for the spin down case $n_1,n_2\in\{1,2,3...\}$ label the Landau levels corresponding to independent magnetic fields in the $(x_1,x_2)$ and $(x_3,x_4)$ planes, and $\ket{n_1,n_2}$ are normalized harmonic oscillator eigenstates. One should keep in mind that (for $B>F$) each level has a degeneracy $M(N-M)$, with $M$ and $(N-M)$ being the fluxes in the $(x_1,x_2)$ and $(x_3,x_4)$ planes. The full spinors are:
\bea
\ket{\psi_\uparrow}&=&\frac{1}{\sqrt{2}}\left( \begin{matrix} \ket{n_1,n_2}\\ 0 \\ -\frac{i}{\lambda}\sqrt{2Fn_2}\,\ket{n_1,n_2-1}\\ -\frac{i}{\lambda}\sqrt{2(B-F)n_1}\,\ket{n_1-1,n_2}\end{matrix}\right)
\nonumber\\
\ket{\psi_\downarrow}&=&\frac{1}{\sqrt{2}}\left( \begin{matrix} 0\\ \ket{n_1-1,n_2-1}\\  -\frac{i}{\lambda}\sqrt{2(B-F)n_1}\, \ket{n_1-1,n_2} \\ -\frac{i}{\lambda}\sqrt{2Fn_2}\,\ket{n_1-1,n_2}\ \end{matrix}\right)
\label{psi+}
\ea
The overall factor fixes the normalization of the spinor to 1. 
In the $\tau_3=-1$ sector, there is a similar construction, recalling that the field strength is $(B+F)$ in the $(x_1,x_2)$ plane.

\section{Matrix Elements and Dipole Moments}

In this Section we consider certain matrix elements involving quark bilinears, such as have been computed on the lattice. For these purposes, it is convenient to introduce a small quark mass $m$, so that the propagator of the
 Dirac operator $\slashed{\mc D}+m$ is given by:
 \bea
 \frac{1}{\slashed{\mc D}+m}=
 \begin{pmatrix}
 \frac{m}{m^2+DD^\dagger} &  \frac{-1}{m^2+DD^\dagger} \, D\cr
 \frac{1}{m^2+D^\dagger D} \, D^\dagger &  \frac{m}{m^2+D^\dagger D}
 \end{pmatrix}
 \eea
 Note that $DD^\dagger$ and $D^\dagger D$ have identical spectra, except for possible zero modes, so they can be viewed as square operators (matrices) of different dimension, as is clear when they are diagonalized in their respective eigenspaces. The zero mode contribution to the propagator can be separated by writing it in one of two ways, depending on which chirality supports zero modes
 \bea
  \frac{1}{\slashed{\mc D}+m}&=&
 \begin{pmatrix}
 \frac{m}{m^2+DD^\dagger} &  \frac{-1}{m^2+DD^\dagger} \, D\cr
 D^\dagger  \frac{1}{m^2+DD^\dagger}& \left(\frac{1}{m}-\frac{1}{m}D^\dagger  \frac{m}{m^2+DD^\dagger} D\right)
 \end{pmatrix} \nonumber\\
 &=&
 \begin{pmatrix}
\left(\frac{1}{m}-\frac{1}{m}D  \frac{m}{m^2+D^\dagger D} D^\dagger\right) &  -D\frac{1}{m^2+D^\dagger D} \cr
   \frac{1}{m^2+D^\dagger D}D^\dagger & \frac{m}{m^2+D^\dagger D}
 \end{pmatrix}
 \eea

 An important set of quark bilinears involve the spin tensor $\Sigma_{\mu\nu}$:
 \bea
 \Sigma_{\mu\nu}&=&\frac{1}{2i}[\gamma_\mu , \gamma_\nu]
=2\begin{pmatrix}
\bar\sigma_{\mu\nu} & 0\cr
0 & \sigma_{\mu\nu}
\end{pmatrix}
\eea
This representation makes clear the natural decomposition of  $\Sigma_{\mu\nu}$ into its self-dual part ($\sigma_{\mu\nu}$) and its anti-self-dual part ($\bar\sigma_{\mu\nu}$). The bilinears are 
\bea
\langle \bar\psi \Sigma_{\mu\nu}\psi\rangle ={\rm tr}\left(\Sigma_{\mu\nu}  \frac{1}{\slashed{\mc D}+m}\right)
\label{bilinear}
\eea
For applications to the chiral magnetic effect, we are interested in the magnetic and electric dipole moments:
\bea
\sigma^{M}_i&=&\frac{1}{2}\epsilon_{ijk}\langle \bar\psi \Sigma_{jk}\psi\rangle 
\eea
\bea\label{dip123}
\sigma^{E}_i&=&\langle \bar\psi \Sigma_{i4}\psi\rangle
\eea
With a strong magnetic field in the $x_3$ direction, we concentrate on $\sigma^{M}_3$ and $\sigma^{E}_3$, which require the spin tensors:
\bea\label{eldip2}
\Sigma_{12}
&=& \begin{pmatrix}
\sigma_{3} & 0\cr
0 & \sigma_{3}
\end{pmatrix}\nonumber\\
\Sigma_{34}
&=& \begin{pmatrix}
-\sigma_{3} & 0\cr
0 & \sigma_{3}
\end{pmatrix}
\eea
Thus,
\bea
m\langle \bar\psi \Sigma_{12}\psi\rangle &=& {\rm tr}_{2\times 2}\left(\sigma_{3}   \frac{m^2}{m^2+DD^\dagger}\right)+{\rm tr}_{2\times 2}\left(\sigma_{3}   \frac{m^2}{m^2+D^\dagger D}\right)\\
m \langle \bar\psi \Sigma_{34}\psi\rangle &=&-{\rm tr}_{2\times 2}\left(\sigma_{3}   \frac{m^2}{m^2+DD^\dagger}\right)+{\rm tr}_{2\times 2}\left(\sigma_{3}   \frac{m^2}{m^2+D^\dagger D}\right)
\label{bilinear2}
\eea
The dominant contribution to the trace over the spectrum comes from the modes with low eigenvalues of $DD^\dagger$ and $D^\dagger D$. 
In the strong magnetic field limit, we see from Figure \ref{fig5} that the zero modes and the near-zero-modes all have spin up, $\sigma_3=+1$, as expected. The dominant contribution to the electric and magnetic moments are therefore:
\bea
m\langle \bar\psi \Sigma_{12}\psi\rangle &\approx &{\rm tr}_{2\times 2}\left(\frac{m^2}{m^2+DD^\dagger}\right)+{\rm tr}_{2\times 2}\left(\frac{m^2}{m^2+D^\dagger D}\right)\\
m \langle \bar\psi \Sigma_{34}\psi\rangle &\approx&-{\rm tr}_{2\times 2}\left( \frac{m^2}{m^2+DD^\dagger}\right)+{\rm tr}_{2\times 2}\left(   \frac{m^2}{m^2+D^\dagger D}\right)
\label{bilinear3}
\eea
For the magnetic dipole moment, the main contribution comes from the zero modes, so we simply count the degeneracies in the various sectors:
\bea
m \langle \bar\psi \Sigma_{12}\psi\rangle &\approx&\left(\frac{B-F}{2\pi}\right)\left(\frac{F}{2\pi}\right)
+\left(\frac{B+F}{2\pi}\right)\left(\frac{F}{2\pi}\right)\nonumber\\
&=&\frac{BF}{2\pi^2}
\label{magnetic}
\eea
which is linear in the magnetic field $B$.
For the electric dipole moment, the near-zero-modes cancel, leaving just  the zero mode contribution:
\bea
m \langle \bar\psi \Sigma_{34}\psi\rangle &\approx&-\left(\frac{B-F}{2\pi}\right)\left(\frac{F}{2\pi}\right)
+\left(\frac{B+F}{2\pi}\right)\left(\frac{F}{2\pi}\right)\nonumber\\
&=&\frac{F^2}{2\pi^2}
\label{electric}
\eea
which is independent of $B$, and negligible compared to $BF$, for $B\gg F$. [Note that (\ref{electric}) does not imply that there is a residual electric dipole moment when $B$ vanishes, because (\ref{electric}) applies only in the $B\gg F$ limit.]
Thus, we see that the zero modes and near-zero-modes imply that
\bea
\langle \bar\psi \Sigma_{12}\psi\rangle\propto B\qquad , \qquad \langle \bar\psi \Sigma_{12}\psi\rangle\gg \langle \bar\psi \Sigma_{34}\psi\rangle
\eea
This is in agreement with the lattice results of \cite{Buividovich:2009my}.

If we now consider the fluctuations in the electric dipole moment, we find a dependence on $B$, because
\bea
\langle \bar\psi \Sigma_{34}\psi\,  \bar\psi \Sigma_{34}\psi\rangle&=&
{\rm tr}\left(  \frac{1}{\slashed{\mc D}+m}\,\Sigma_{34}\, \frac{1}{\slashed{\mc D}+m}\,\Sigma_{34}\right)\nonumber\\
&=&{\rm tr}_{2\times 2}\left(\frac{m^2}{(m^2+DD^\dagger)^2}+\frac{1}{(m^2+DD^\dagger)}\,D\sigma_3D^\dagger\sigma_3 \frac{1}{(m^2+DD^\dagger)}\right)+\nonumber\\
&&{\rm tr}_{2\times 2}\left(\frac{1}{(m^2+D^\dagger D)^2}D^\dagger\sigma_3D\sigma_3+\frac{m^2}{(m^2+D^\dagger D)}\sigma_3  \frac{1}{(m^2+D^\dagger D)}\sigma_3\right)\nonumber\\
&\approx& {\rm tr}_{2\times 2}\left(\frac{1}{(m^2+DD^\dagger)}+\frac{1}{(m^2+D^\dagger D)}\right)
\eea
where in the last step we have used the fact that the dominant contribution comes from zero modes and near-zero-modes, all of which have $\sigma_3=+1$. Thus, comparing with (\ref{magnetic}) we see that the fluctuation is linear in $B$
\bea
\langle \bar\psi \Sigma_{34}\psi\,  \bar\psi \Sigma_{34}\psi\rangle \approx \left(\frac{ F}{2\pi^2 m^2 L^4}\right)\, B
\label{electric-fluc}
\eea
again in agreement with the lattice results of \cite{Buividovich:2009my}.

\section{Small instanton limit}

In the opposite limit of a weak magnetic field the radius of the Landau orbit is much larger than the instanton size:  $1/\sqrt{B} \gg \rho$. It appears that a quantitative analysis is more difficult in this case, because all Landau levels contribute. However, we can still outline the qualitative picture.  The effects  induced by the presence of a small instanton on quark dynamics can be described in terms of the effective lagrangian introduced by 't Hooft \cite{'tHooft:1976fv}:
\bea\label{thooft}
{\cal L}(x) = \kappa\ e^{i \theta}\det \left[-\bar{\psi}_R(x) \psi_L(x)\right] + h.c.,
\eea
where $\kappa$ is a constant that contains $\exp(-8 \pi^2/g^2)$, $\theta$ is the $\theta$-angle of QCD (that we will assume be equal to zero), and the subscripts $L$ and $R$ refer to the left- and right-handed quark helicities. The flavor determinant breaks the $U_L(N_f) \times U_R(N_f)$ symmetry of QCD down to $SU_L(N_f) \times SU_R(N_f)$. The (anti)instanton vertex as described by (\ref{thooft}) absorbs $N_f$ left-handed fermions and creates the same number of right-handed ones, or vice versa. 

Let us now turn on an external magnetic field directed along $x_3$, and consider the electric dipole moment given by (\ref{dip123}). The Wigner-Eckart theorem tells us that the only possible orientation of the dipole moment is along the $x_3$ axis. Using the chiral representation of the quark spinors as in (\ref{thooft}), and using the gamma matrices (\ref{eldip2}), we can write down the electric dipole moment as
\bea\label{eldipes}
\sigma_3^E = - {\bar{\psi}}_L \sigma_3 \psi_L + {\bar{\psi}}_R \sigma_3 \psi_R\quad  .
\eea
This expression makes it clear that in the absence of an asymmetry between the left- and right-handed fermions the electric dipole moment should vanish identically, as required by P and CP invariances. However, the (anti)instanton transition caused by the interaction (\ref{thooft}) can create a local asymmetry between the left- and right-handed fermions, and thereby induce a non-zero local electric dipole moment.

It is interesting to discuss further the physical  origin of this effect. In the absence of an instanton,  in a magnetic field and with a fixed spin projection $\sigma_3=+1$, there exist an equal number of left- and right-handed zero modes, as we discussed above in Section \ref{magback}. As a result, they cancel each other in (\ref{eldipes}). However, the instanton induces  couplings between the fermion's spin, isospin (that belongs to the $SU(2)$ subgroup of the color group $SU(N_c)$) and the (four-dimensional) orbital angular momentum, so that the operator $D^2$ contains the isospin-orbit term $\sim \vec{T} \cdot \vec{L}$ \cite{'tHooft:1976fv}. Because of this, the angular momentum of the fermion ceases to be a conserved quantum number -- only a combination $\vec{J} = \vec{L} + \vec{S} + \vec{T}$ is conserved. The presence of the isospin-orbit term leads to the mixing of the lowest Landau level of left chirality with a radial excitation of right chirality that has the opposite parity. The change of the orbital angular momentum is compensated by a rotation in the color $SU(2)$ sub-space. Because of this, the system acquires an electric dipole moment $\sigma_3^E \sim \kappa$ signaling a local violation of parity invariance. 
The emergence of a local quark  electric dipole moment in an external magnetic field has been observed on the lattice in \cite{Buividovich:2009my}, where it was also observed that the electric dipole moment is strongly correlated with the local chiral density signaling the presence of the instanton, or some other topological object.
Of course, at $\theta=0$ the action (\ref{thooft}) does not generate an asymmetry between the instantons and anti-instantons, so there is no global violation of parity. 

\section{Conclusions}

While the physics of fermions in magnetic fields and instantons, separately, is well known and well understood, we have shown here that the combination of both background fields leads to a surprisingly intricate and rich structure in the Dirac spectrum. The inherent asymmetry when both instanton and magnetic field are present can lead to the development of an electric dipole moment. Physically, it can be understood as the outcome of two competing effects: the spin projection produced by a magnetic field and the chirality projection produced by an instanton field. We have illustrated this in detail both in the strong magnetic field limit in which the instanton scale is large compared to the magnetic length, and also in the opposite limit of small instantons. We have used the language of a four dimensional torus in Euclidean space, motivated in part by a desire to connect analytic results with recent lattice studies, which have shown  a wide variety of interesting effects arising from the coupling of QCD to electromagnetic fields \cite{Buividovich:2009wi,Buividovich:2009my,Abramczyk:2009gb,tom,Tiburzi:2011vk}. Corrections to the large instanton limit case could be constructed using the derivative expansion, in the natural Fock-Schwinger gauge \cite{Cronstrom:1980hj}, $x_\mu \mc A_\mu=0$, which can be chosen for the combined instanton-magnetic field background, and which is an efficient means for computing induced currents, expectation values and correlators \cite{Shifman:1980ui,Dubovikov:1981bf,Ioffe:1983ju}. While in a constant self-dual field there is no preferred position at which the modes are localized, in the next order of the Fock-Schwinger gauge expansion we expect that these zero modes and near-zero-modes will localize on the instantons, as in the phenomenon of dynamical localization in the quantum Hall effect \cite{arovas}.

\bigskip
We thank T. Blum,  M. Polikarpov and E. Shuryak for helpful discussions.
This work was supported by the US Department of Energy under grants DE-FG02-92ER40716 (GB and GD) and DE-AC02-98CH10886, DE-FG-88ER41723 (GB and DK). 

\end{document}